\documentclass[12pt,prd,tightenlines,preprintnumbers,showpacs,nofootinbib,
floatfix,superscriptaddress]{revtex4}
\usepackage{epsfig}
\usepackage{amssymb,amsmath}
\usepackage{color,epsfig,graphicx,longtable}

\newcommand{\Dlr}{\overset{\leftrightarrow}{D}}
\newcommand{\gRcf}{\frac{g_R^2 \, C_F}{16 \pi^2} }
\newcommand{\gTIcf}{\frac{g_{TI}^2 \, C_F}{16 \pi^2} }
\newcommand{\gR}{g_{\mathrm R}}
\newcommand{\cO}{\mathcal {O}}
\newcommand{\MS}{{\overline{\mbox{MS}}}}

\begin{document}

\preprint{DESY 06-050} 
\preprint{Edinburgh 2006/08} 
\preprint{Liverpool LTH 698}
\preprint{LU-ITP 2006/005} 

\title{Renormalisation of Composite Operators in Lattice Perturbation Theory
with Clover Fermions: Non-forward Matrix Elements}

\author{M.~G\"ockeler}\affiliation{Institut f\"ur Theoretische Physik, 
        Universit\"at Regensburg, 93040 Regensburg, Germany}
\author{R.~Horsley}\affiliation{School of Physics, University of 
        Edinburgh, Edinburgh EH9 3JZ, UK}
\author{H.~Perlt}\affiliation{Institut f\"ur Theoretische Physik, 
        Universit\"at Leipzig, 04109 Leipzig, Germany}
\author{P.E.L.~Rakow}\affiliation{Theoretical Physics Division, 
       Department of Mathematical Sciences, University of Liverpool, 
        Liverpool L69 3BX, UK}
\author{A.~Sch\"afer}\affiliation{Institut f\"ur Theoretische Physik, 
        Universit\"at Regensburg, 93040 Regensburg, Germany}
\author{G.~Schierholz}\affiliation{John von Neumann-Institut f\"ur 
        Computing NIC, Deutsches Elektronen-Synchrotron DESY, 
        15738 Zeuthen, Germany}
       	\affiliation{Deutsches Elektronen-Synchrotron DESY, 
        22603 Hamburg, Germany}
\author{A.~Schiller}\affiliation{Institut f\"ur Theoretische Physik, 
        Universit\"at Leipzig, 04109 Leipzig, Germany}

\begin{abstract}
We consider the renormalisation of lattice QCD operators with 
one and two covariant derivatives related to the first and second 
moments of generalised parton distributions and meson 
distribution amplitudes. Employing the clover fermion action we 
calculate their non-forward quark matrix elements in one-loop 
lattice perturbation theory.
For some representations of the hypercubic group commonly used in 
simulations we determine the sets of all possible mixing operators 
and compute the matrices of renormalisation factors 
in one-loop approximation.
We describe how tadpole improvement is applied to the results.
\end{abstract}


\date{}

\maketitle

\section{Introduction}

Many interesting observables in hadron physics, e.g.\ (moments of)
generalised parton distributions (GPDs)~\cite{gpd} or distribution amplitudes, 
can be computed from matrix elements of local operators between hadron states. 
(For an extensive review of GPDs see Ref.~\cite{Diehl}, for distribution
amplitudes see, e.g., Ref.~\cite{DA}.)
GPDs, in particular, have attracted a lot of interest in recent years.
They parametrise a 
large class of hadronic correlators, including e.g.\ form factors
and the ordinary parton distribution functions. Thus GPDs provide 
a formal framework to connect information from various inclusive, 
semi-inclusive and exclusive reactions. 
Furthermore they give access to physical quantities which 
cannot be directly determined in experiments, like e.g.\ the orbital 
angular momentum of quarks and gluons in a nucleon (in a given
scheme) and the spatial distribution of the energy or spin 
density of a fast moving hadron in the transverse plane.
On the other hand, direct experimental information is limited
and additional input is required to obtain a more complete knowledge
of GPDs. One important source is lattice QCD, which
can provide the relevant hadronic matrix 
elements~\cite{Hagler:2003jd,QCDSF,Gockeler:2004vx,Gockeler:2005aw,flip}.

Compared with moments of ordinary parton distributions, the specific
difficulty in the treatment of moments of GPDs and distribution
amplitudes lies in the fact that the required matrix elements of
local operators are no longer forward matrix elements.
In general, this circumstance complicates the pattern of mixing under
renormalisation, and more extended investigations become necessary.

In a recent publication~\cite{Gockeler:2004xb} we have calculated the
non-forward quark matrix elements needed for the renormalisation of 
quark-antiquark operators with two derivatives, which determine 
the second moments of GPDs, and we have discussed the mixing problem
in detail. This calculation was performed in one-loop lattice 
perturbation theory for the Wilson fermion action.

In the present paper we extend our perturbative calculations to 
improved fermions using the Sheikholeslami-Wohlert (clover)
action~\cite{Sheikholeslami:1985ij} for improving the vertices. 
We calculate renormalisation factors, but no improvement coefficients.
While the general framework is of course the
same as for Wilson fermions, the additional clover term in the action 
leads to a considerable complication of the calculations.
A preliminary account of our work has
already been given in Ref.~\cite{cyprus}. Note, however, that a few
misprints in Ref.~\cite{cyprus} will be corrected here.

Let us fix the notations used in our perturbative calculations.
We work in Euclidean space and employ the Wilson gauge action together with
clover fermions such that the total action is given by
\begin{equation}
  S^{\rm latt} = S_{\rm SW,F} +S_{\rm W,G}\,.
\end{equation}
The fermionic part $S_{\rm SW,F}$
has the form~\cite{Sheikholeslami:1985ij}
\begin{eqnarray}
  S_{\rm SW,F}&=&
  4 r a^3 \sum_{x} \bar{\psi}(x)\psi(x)
  \nonumber
  \\
  &-& \frac{a^3}{2}
  \sum_{x,\mu}\left[\bar{\psi}(x)(r-\gamma_\mu)U_{x,\mu} \psi(x+a \hat{\mu})
  + \bar{\psi}(x+a \hat{\mu})(r+\gamma_\mu)U^\dagger_{x,\mu} \psi(x)\right]
  \\
  &-& \mathrm i \frac{a^5\,g\,c_{sw}}{8}
  \sum_{x,\mu,\nu}\,\bar{\psi}(x) [\gamma_\mu , \gamma_\nu ]
                F^{\rm clover}_{\mu\nu} \psi(x)
  \nonumber
\nonumber
\end{eqnarray}
written in terms of dimensionful massless fermion fields $\psi(x)$. 
Here $a$ denotes  the lattice spacing and the sums run over all lattice
sites $x$ and directions $\mu,\nu$. All other indices are suppressed.
$F^{\rm clover}_{\mu\nu}$ is the standard ``clover-leaf'' form of the
lattice field strength (see, e.g., Appendix D in Ref.~\cite{timid}).
The coupling strength of the improvement term is given by $c_{sw}$.
The link matrices $U_{x,\mu}$ are related to the gauge field $A_\mu(x)$ by
\begin{equation}
  U_{x,\mu} = \exp \left[\mathrm i g a A_\mu(x)\right],  \quad 
                                 A_\mu(x)=T^c A^c_\mu(x)\,,
\end{equation}
where $g$ is the bare gauge coupling and the $T^c$ are
the generators of the $SU(3)$ algebra.
The gauge action for the gluon field $A_\mu(x)$ is
\begin{equation}
  S_{\rm W,G} = \frac{6}{g^2} \sum _{x,\mu<\nu}\left[1 -
  \frac{1}{6}{\rm Tr}\left(U_{x,\mu\nu}+U^\dagger_{x,\mu\nu}\right)\right]
\end{equation}
with
\begin{equation}
  U_{x,\mu\nu} = U_{x,\mu}U_{x+a\hat{\mu},\nu}U^\dagger_{x+a\hat{\nu},\mu}
  U^\dagger_{x,\nu}
  \,.
\end{equation}

In a perturbative calculation the investigated operators are
considered between off-shell quark states.
Our calculations are performed in Feynman gauge, the final numbers
will be presented for the Wilson parameter $r=1$, but for arbitrary
values of $c_{sw}$.

On the lattice the operators are classified according to the
irreducible representations $\tau_k^{(l)}$ of the hypercubic group 
H(4) (for the notation see, e.g., Ref.~\cite{group}).
Here $l$ denotes the dimension of the representation and $k$ 
labels inequivalent representations of the same dimension. In addition 
our operators will be chosen such that they have 
definite charge conjugation parity $C$.

Using clover fermions, quark-antiquark operators with one covariant 
derivative have been discussed 
in~\cite{Capitani:2000xi} for forward matrix elements. 
Because of the constraints imposed by charge conjugation invariance
no additional mixing has to be considered in non-forward matrix elements 
and the renormalisation constants given there can be taken over to the 
case at hand. We will present them again for completeness and add 
the corresponding results for operators involving $[\gamma_\mu, \gamma_\nu]$, 
which were not considered in Ref.~\cite{Capitani:2000xi}.

In~\cite{Gockeler:2004xb} the renormalisation procedure for the
case of non-vanishing momentum transfer has been discussed in detail. 
We will not repeat this discussion here. The matrix of 
renormalisation and mixing coefficients $Z_{ij}(a\mu)$ relating 
regularised lattice vertex functions $\Gamma_j^L (p',p,a,\gR)$ 
and $\MS$ renormalised vertex functions $\Gamma_i^R (p',p,\mu,\gR)$ 
is defined such that
\begin{equation}
  \Gamma_i^R (p',p,\mu,\gR) = Z_\psi^{-1}(a\mu)
  \sum_{j=1}^N Z_{ij}(a\mu) \, \Gamma_j^L (p',p,a,\gR)
  \label{renorm}
\end{equation}
with the quark wave function renormalisation constant
$Z_\psi$. Here $p$ ($p'$) denotes the momentum of the incoming 
(outgoing) quark, the renormalisation scale is $\mu$, the 
renormalised coupling is denoted by $\gR$, and $N$ is the number 
of operators which mix in the one-loop approximation.

\section{Operators and mixing}
\label{OpMix}

We consider operators with up to two covariant symmetric lattice derivatives
$\overset{\leftrightarrow}{D} = \overset{\to}{D} 
 - \overset{\leftarrow}{D} $ 
and external ordinary derivatives $\partial$. 
In the operator symbols, the derivatives are indicated by superscripts 
$D$ and $\partial$. Note that matrix elements of operators which are
ordinary derivatives of other operators are simply given by the 
matrix elements of these other operators multiplied by the appropriate
product of components of the momentum transfer.

The quark-antiquark operators with one derivative are given by
\begin{eqnarray}
 \cO^D_{\mu\nu}&=& -\frac{\mathrm i}{2}
  \bar\psi \gamma_\mu \Dlr_\nu \psi\,,
  \label{O11} \\
  \cO^{5,D}_{\mu\nu}&=& -\frac{\mathrm i}{2}
  \bar\psi \gamma_\mu \gamma_5\Dlr_\nu \psi\,,
  \label{O12} \\
  \cO^{T,D}_{\mu\nu\omega}&=&-\frac{\mathrm i}{2}
  \bar\psi [ \gamma_\mu , \gamma_\nu ] \Dlr_\omega \psi \,,
  \label{Oplow1} \\
  \cO^{T,\partial}_{\mu\nu\omega}&=&-\frac{\mathrm i}{2}
  \partial_\omega
  \left( \bar\psi  [\gamma_\mu,\gamma_\nu ]  \psi \right)\,.
  \label{Oplow2}
\end{eqnarray}
Operators such as (\ref{Oplow1}) involving $[\gamma_\mu,\gamma_\nu]$
are of interest for tensor GPDs as well as for transversity and we call them
transversity operators. They are antisymmetric in the indices $\mu$ 
and $\nu$. For non-chiral fermions, operators 
(\ref{Oplow1}) and (\ref{Oplow2}) contribute as lower-dimensional operators
to mixing in certain operators which determine second moments of GPDs. 

As operators with two derivatives we consider
\begin{eqnarray}
  \cO_{\mu\nu\omega}^{DD}&=& -\frac{1}{4}
  \bar\psi \gamma_\mu \Dlr_\nu \Dlr_\omega\psi\,,
  \nonumber \\
  \cO_{\mu\nu\omega}^{\partial D}&=& -\frac{1}{4}
  \partial_\nu \left( \bar\psi \gamma_\mu \Dlr_\omega\psi \right) \,,
  \label{OpDD}
  \\
  \cO_{\mu\nu\omega}^{\partial \partial}&=& -\frac{1}{4}
  \partial_\nu \partial_\omega
  \left(\bar\psi \gamma_\mu \psi \right)
  \nonumber
\end{eqnarray}
and
\begin{eqnarray}
  \cO_{\mu\nu\omega}^{5,DD}&=&-\frac{1}{4}
  \bar\psi \gamma_\mu \gamma_5 \Dlr_\nu \Dlr_\omega\psi\,,
  \nonumber
  \\
  \cO_{\mu\nu\omega}^{5,\partial D}&=& -\frac{1}{4}
  \partial_\nu \left( \bar\psi \gamma_\mu \gamma_5 \Dlr_\omega \psi\right) \,,
  \label{Op5DD}
  \\
  \cO_{\mu\nu\omega}^{5,\partial \partial}&=&-\frac{1}{4}
  \partial_\nu \partial_\omega
  \left(\bar\psi \gamma_\mu \gamma_5\psi \right) \,.
  \nonumber
\end{eqnarray}
We include here also transversity operators with two derivatives:
\begin{equation}
  \cO_{\mu\nu\omega\sigma}^{T,DD}=-\frac{1}{4}\bar\psi [\gamma_\mu,\gamma_\nu]
  \Dlr_\omega   \Dlr_\sigma\psi\,,
  \quad
  \cO_{\mu\nu\omega\sigma}^{T,\partial\partial }=
  -\frac{1}{4} \partial_\omega\partial_\sigma
  \left( \bar\psi [\gamma_\mu,\gamma_\nu] \psi  \right) \,.
  \label{Optrans}
\end{equation}
A detailed description of the operators with two derivatives,
their representations and potential mixings
is given in~\cite{Gockeler:2004xb}.
To define the various operators we use the following short-hand notations:
\begin{eqnarray}
  \cO_{\cdots\{ \nu_1 \nu_2 \} }&=&
    \frac{1}{2} \left( \cO_{\cdots \nu_1\nu_2}+\cO_{\cdots \nu_2\nu_1} \right)
  \,,
  \\
  \cO_{ \{ \nu_1\nu_2\nu_3 \} }&=& \frac{1}{6} \left(
  \cO_{\nu_1\nu_2\nu_3}+\cO_{\nu_1\nu_3\nu_2}+\cO_{\nu_2\nu_1\nu_3} +
  \cO_{\nu_2\nu_3\nu_1}+\cO_{\nu_3\nu_1\nu_2}+\cO_{\nu_3\nu_2\nu_1}  \right)
  \,,
  \\
  \cO_{\|\nu_1\nu_2\nu_3\| } &=& \cO_{\nu_1\nu_2\nu_3}-\cO_{\nu_1\nu_3\nu_2}+
  \cO_{\nu_3\nu_1\nu_2}-\cO_{\nu_3\nu_2\nu_1}-2\,\cO_{\nu_2\nu_3\nu_1}
  +2\,\cO_{\nu_2\nu_1\nu_3}
  \,,
  \\
  \cO_{\langle\langle\nu_1\nu_2\nu_3\rangle\rangle } &=&
  \cO_{\nu_1\nu_2\nu_3}+\cO_{\nu_1\nu_3\nu_2}
  -\cO_{\nu_3\nu_1\nu_2}-\cO_{\nu_3\nu_2\nu_1}
  \,.
\end{eqnarray}

For the first moments we choose the following representations and operators
(for a more detailed
discussion of the transformation under H(4) see~\cite{group}):
\begin{equation}
\begin{tabular}{lcc}
  Operator  & Representation & \mbox{$C$} \\
  \hline\\[-2ex]
  $\cO^D_{\{14\}}$                                                  
    & $\tau_3^{(6)}$ & $+1$\\ [0.7ex]
  $\cO^D_{44}-\frac{1}{3}\left(\cO^D_{11}+\cO^D_{22}+\cO^D_{33}\right) $    
    & $\tau_1^{(3)}$ & $+1$\\ [0.7ex]
  $\cO^{5,D}_{\{14\}}$ 
    & $\tau_4^{(6)}$ & $-1$\\ [0.7ex]
  $\cO^{5,D}_{44}-\frac{1}{3}\left(\cO^{5,D}_{11}+\cO^{5,D}_{22}
                                           +\cO^{5,D}_{33}\right) $ 
    & $\tau_4^{(3)}$ & $-1$\\ [0.7ex]
  $\cO^{T,D}_{\langle\langle124\rangle\rangle}$ 
    & $\tau_2^{(8)}$ & $+1$\\ [0.7ex]
  $\cO^{T,D}_{\langle\langle122\rangle\rangle}-
                         \cO^{T,D}_{\langle\langle133\rangle\rangle}$ 
    & $\tau_1^{(8)}$ & $+1$\\ [0.7ex]
 \end{tabular}
\vspace*{0.5cm}
\label{tabO}
\end{equation}
All operators in (\ref{tabO}) are multiplicatively renormalisable.
These representations exhaust all possibilities for the twist-2 sector
in the continuum. Note that in Ref.~\cite{cyprus} we have erroneously
assigned $\tau_1^{(8)}$ and $C=-1$ to an operator belonging to 
$\tau_1^{(8)}$, $C=+1$.

Let us now turn to the second moments and the corresponding twist-2
operators. In the unpolarised case we consider the 
following sets of mixing operators:

\vspace{0.5cm}
\underline{$\tau_2^{(4)},C=-1$}
\begin{equation}
  \cO_{\{124\}}^{DD} \,, \, \cO_{\{124\}}^{\partial\partial} \,,
  \label{mixing1}
\end{equation}

\underline{$\tau_1^{(8)},C=-1$}
\begin{eqnarray}
 \label{mixing2}
 &&\cO_1=\cO^{DD}_{\{114\}}-\frac{1}{2}
  \left(\cO^{DD}_{\{224\}}+\cO^{DD}_{\{334\}}\right)
  \,,
  \nonumber
  \\
  &&\cO_2=\cO^{\partial\partial}_{\{114\}}-\frac{1}{2} \left(
   \cO^{\partial\partial}_{\{224\}}
  +\cO^{\partial\partial}_{\{334\}}\right)
  \,,
  \nonumber
  \\
  &&\cO_3=\cO^{DD}_{\langle\langle 114\rangle\rangle}-\frac{1}{2}
  \left( \cO^{DD}_{\langle\langle224\rangle\rangle}+
         \cO^{DD}_{\langle\langle334\rangle\rangle}\right)
  \,,
  \nonumber
  \\
  &&\cO_4=\cO^{\partial\partial}_{\langle\langle 114 \rangle\rangle}
  -\frac{1}{2}
     \left(  \cO^{\partial\partial}_{\langle\langle 224 \rangle\rangle}+
          \cO^{\partial\partial}_{\langle\langle 334 \rangle\rangle}\right)
  \,,
  \label{O4}
  \\
  &&\cO_5=\cO^{5,\partial D}_{||213||}
  \,,
  \nonumber
  \\
  &&\cO_6=\cO^{5,\partial D}_{\langle\langle213\rangle\rangle}
  \,,
  \nonumber
  \\
  &&\cO_7=\cO^{5,DD}_{||213||}
  \,,
  \nonumber
  \\
  &&\cO_8=   \cO^{T,\partial}_{411}-
  \frac{1}{2}\left(\cO^{T,\partial}_{422}+\cO^{T,\partial}_{433} \right)
  \,.
  \nonumber
\end{eqnarray}

There is one more representation, $\tau_1^{(4)},C=-1$, giving twist-2
operators. However, even in forward matrix elements the corresponding
operators mix with operators whose dimension is smaller by two. Therefore
they are rather unsuitable for numerical simulations and will not be 
discussed any further.

In the polarised case we consider

\underline{$\tau_3^{(4)},C=+1$}
\begin{equation}
  \label{mixing3}
  \cO_{\{124\}}^{5,DD} \, , \, \cO_{\{124\}}^{5,\partial\partial} \,,
\end{equation}

\underline{$\tau_2^{(8)},C=+1$}
\begin{eqnarray}
  \label{mixing4}
  &&\cO^5_1=\cO^{5,DD}_{\{114\}}-\frac{1}{2}
  \left(\cO^{5,DD}_{\{224\}}+\cO^{5,DD}_{\{334\}}\right)
  \,,
  \nonumber
  \\
  &&\cO^5_2=\cO^{5,\partial\partial}_{\{114\}}-\frac{1}{2} \left(
  \cO^{5,\partial\partial}_{\{224\}}
  +\cO^{5,\partial\partial}_{\{334\}}\right)
  \,,
  \nonumber
  \\
  &&\cO^5_3=\cO^{5,DD}_{\langle\langle 114\rangle\rangle}-\frac{1}{2}
  \left( \cO^{5,DD}_{\langle\langle224\rangle\rangle}+
         \cO^{5,DD}_{\langle\langle334\rangle\rangle}\right)
  \,,
  \nonumber
  \\
  &&\cO^5_4=\cO^{5,\partial\partial}_{\langle\langle 114
  \rangle\rangle}-\frac{1}{2}
  \left(  \cO^{5,\partial\partial}_{\langle\langle 224 \rangle\rangle}+
       \cO^{5,\partial\partial}_{\langle\langle 334 \rangle\rangle}\right)
  \,,
  \label{O54}
  \\
  &&\cO^5_5=\cO^{\partial D}_{||213||}
  \,,
  \nonumber
  \\
  &&\cO^5_6=\cO^{\partial D}_{\langle\langle213\rangle\rangle}
  \,,
  \nonumber
  \\
  &&\cO^5_7=\cO^{DD}_{||213||}
  \,,
  \nonumber
  \\
  &&\cO^5_8 =
  \cO^{T,D}_{123} -  2\cO^{T,D}_{231} - \cO^{T,D}_{132} \,.
  \nonumber
\end{eqnarray}

Again, there is one more representation giving twist-2 operators. 
Here it is $\tau_4^{(4)},C=+1$, but the corresponding operators 
suffer from similar mixing problems as the operators with 
$\tau_1^{(4)},C=-1$ in the unpolarised case and will be omitted in the 
following.

Finally the transversity operators:

\underline{$\tau_2^{(3)},C=-1$}
\begin{equation}
  \label{mixing5}
  \cO^T_1=\cO^{T,DD}_{4\{123\}} \, , \, 
    \cO^T_2=\cO^{T,\partial\partial}_{4\{123\}} \,,
\end{equation}

\underline{$\tau_3^{(3)},C=-1$}
\begin{eqnarray}
  \label{mixing6}
  &&\cO^T_3=-\cO^{T,DD}_{1\{133\}}+\cO^{T,DD}_{1\{144\}}
          -\cO^{T,DD}_{2\{233\}}+\cO^{T,DD}_{2\{244\}}
         -2\cO^{T,DD}_{3\{344\}}
  \,,
  \nonumber
  \\
  &&\cO^T_4=-\cO^{T,\partial\partial}_{1\{133\}}
          +\cO^{T,\partial\partial}_{1\{144\}}
          -\cO^{T,\partial\partial}_{2\{233\}}
          +\cO^{T,\partial\partial}_{2\{244\}}
         -2\cO^{T,\partial\partial}_{3\{344\}} \,,
\end{eqnarray}

\underline{$\tau_2^{(6)},C=-1$}
\begin{eqnarray}
  \label{mixing7}
  &&\cO^T_5 = \cO^{T,DD}_{13\{32\}} + \cO^{T,DD}_{23\{31\}}
          - \cO^{T,DD}_{14\{42\}} - \cO^{T,DD}_{24\{41\}}
  \,,
  \nonumber
  \\
  &&\cO^T_6 =  \cO^{T,\partial\partial}_{13\{32\}}
           + \cO^{T,\partial\partial}_{23\{31\}}
           - \cO^{T,\partial\partial}_{14\{42\}}
           - \cO^{T,\partial\partial}_{24\{41\}} \,.
\end{eqnarray}

The representations (\ref{mixing5}), (\ref{mixing6}) and (\ref{mixing7})
exhaust all possibilities for transversity operators of twist 2
which have a mixing matrix of
size $2 \times 2$ only. Other representations have more complicated 
mixing patterns.

\section{One-loop calculation}

We calculate the matrix elements of the operators in one-loop
lattice perturbation theory in the infinite volume limit
following Kawai et al.~\cite{Kawai:1980ja}.
Details of the computational procedure, in particular the Feynman 
rules, are given in~\cite{Gockeler:2004xb}.

\subsection{First Moment}

Since mixing is absent we need no matrix indices
for the renormalisation constants and use the general form
\begin{equation}
  Z(a\mu) = 1 -
  \gRcf \left(\gamma\,\ln(a^2\mu^2) + B(c_{sw})\right)\,,
  \label{Z}
\end{equation}
where the finite piece $B$ depends on $c_{sw}$ and the (one-loop)
anomalous dimension $\gamma$ is given by
\begin{equation}
\gamma = \left\{
        \begin{array}{ll}
        $8/3$          & \textrm{for   $\quad \tau_3^{(6)}\,,
                      \tau_1^{(3)}\,,\tau_4^{(6)}\,,\tau_4^{(3)}$} \\[0.7ex]
	$3$            & \textrm{for   $\quad \tau_2^{(8)}\,,\tau_1^{(8)} $}
	\end{array}
	\right. \,.
\end{equation}
For the operators (\ref{tabO}) we have (results for $\tau_3^{(6)}$,  
$\tau_1^{(3)}$, $\tau_4^{(6)}$ and $\tau_4^{(3)}$ are taken 
from~\cite{Capitani:2000xi})
\begin{equation}
\begin{tabular}{cr@{.}l}
  Representation & \multicolumn{2}{c}{$B(c_{sw})$} \\
  \hline\\[-2ex]
  $\tau_3^{(6)}$ & 1 & $ 27959 - 3.87297\,c_{sw} -0.67826\,c_{sw}^2 $\\ [0.7ex]
  $\tau_1^{(3)}$ & 2 & $ 56184 - 3.96980\,c_{sw} -1.03973\,c_{sw}^2 $\\ [0.7ex]
  $\tau_4^{(6)}$ & 0 & $ 34512 - 1.35931\,c_{sw} -1.89255\,c_{sw}^2 $\\ [0.7ex]
  $\tau_4^{(3)}$ & 0 & $ 16738 - 1.24953\,c_{sw} -1.99804\,c_{sw}^2 $\\ [0.7ex]
  $\tau_2^{(8)}$ & 1 & $ 25245 - 3.10180\,c_{sw} -1.59023\,c_{sw}^2 $\\ [0.7ex]
  $\tau_1^{(8)}$ & 0 & $ 52246 - 2.99849\,c_{sw} -1.46224\,c_{sw}^2 $
  \end{tabular}
\vspace*{0.5cm}
\label{tabZ}
\end{equation}

\subsection{Second Moment}

We write the matrix of renormalisation constants in the generic form
\begin{equation}
  Z_{ij}^{(m)}(a\mu) = \delta_{ij} -
  \gRcf \left(\gamma_{ij}\,\ln(a^2\mu^2)
  + B_{ij}^{(m)}(c_{sw}) \right)\,,
  \label{Zjk}
\end{equation}
with~\footnote{ $B_{ij}^{(0,m)}$ has been denoted by $c_{ij}^{(m)}$ 
in \cite{Gockeler:2004xb}.}
\begin{equation}
B_{ij}^{(m)}(c_{sw})=B_{ij}^{(0,m)}+B_{ij}^{(1,m)}\,c_{sw}+
		B_{ij}^{(2,m)}\,c_{sw}^2\,.
\end{equation}
The superscript $(m)$ with $m=I,II$ distinguishes the
realisations I and II of the covariant derivatives, which are explained
in Appendix A of~\cite{Gockeler:2004xb}. In the first case the momentum
transfer ``acts'' at the position $x$ associated with the operator, 
where we define for one covariant derivative
\begin{eqnarray}
\cO (x) =  \left(\bar{\psi}\Dlr_\mu \psi\right) (x)
 &=& \frac{1}{2a} \Big( \bar{\psi}(x)U_{x,\mu} \psi(x+a\hat{\mu})
 - \bar{\psi}(x)U^\dagger_{x-a\hat{\mu},\mu} \psi(x-a\hat{\mu}) 
\nonumber \\ & & \qquad {}
 + \bar{\psi}(x-a\hat{\mu})U_{x-a\hat{\mu},\mu} \psi(x)
 - \bar{\psi}(x+a\hat{\mu})U^\dagger_{x,\mu} \psi(x)  \Big) \,.
\end{eqnarray}
We have set the Dirac matrix in the operator equal to the unit 
matrix for simplicity. Realisation I leads to
\begin{eqnarray}
 \left(\bar{\psi}\Dlr_\mu \psi\right)^{(I)}\!\!(q) =
  \frac{1}{2a}\sum_x \left[\bar{\psi}(x)U_{x,\mu} \psi(x+a\hat{\mu})-
  \bar{\psi}(x+a\hat{\mu})U^\dagger_{x,\mu} \psi(x)\right]\,
   \left[{\rm e}^{\mathrm i q\cdot  x}
          +{\rm e}^{\mathrm i q \cdot(x+a\hat{\mu})}\right]\,.
  \label{DI}
\end{eqnarray}
Alternatively (realisation II), $q$ can be 
applied at the point half way between the $\bar{\psi}$ and $\psi$ fields:
\begin{equation}
  \left(\bar{\psi}\Dlr_\mu \psi\right)^{(II)}\!\!(q) =
  \frac{1}{a}\sum_x \left[\bar{\psi}(x)U_{x,\mu} \psi(x+a\hat{\mu})-
  \bar{\psi}(x+a\hat{\mu})U^\dagger_{x,\mu} \psi(x)\right]\,
  {\rm e}^{\mathrm i q   \cdot(x+a\hat{\mu}/2)}\, .
  \label{DII}
\end{equation}
In the case of two covariant derivatives we have for realisation II
\begin{eqnarray}
\left(\bar{\psi}\Dlr_\mu\Dlr_\nu \psi\right)^{(II)}(q)&=& 
\frac{1}{a^2}\sum_x \big(\bar{\psi}(x)U_{x,\mu}U_{x+a\hat{\mu},\nu}
   \psi(x+a\hat{\mu}+a\hat{\nu})
\nonumber\\
& & \quad -\bar{\psi}(x+a\hat{\nu})U_{x+a\hat{\nu},\mu}
                     U^\dagger_{x+a\hat{\mu},\nu}\psi(x+a\hat{\mu})
\nonumber\\
& & \quad -\bar{\psi}(x+a\hat{\mu})U^\dagger_{x,\mu}U_{x,\nu}
                                               \psi(x+a\hat{\nu})
\nonumber\\
& & \quad +\bar{\psi}(x+a\hat{\mu}+a\hat{\nu})
    U^\dagger_{x+a\hat{\nu},\mu}U^\dagger_{x,\nu}\psi(x)\big)\, 
        \mathrm{e}^{\mathrm{i}q\cdot(x+a\hat{\mu}/2+a\hat{\nu}/2)} \,,
\label{DDII}
\end{eqnarray}
and realisation I is obtained from (\ref{DDII}) as
\begin{equation}
\left(\bar{\psi}\Dlr_\mu\Dlr_\nu \psi\right)^{(I)}(q) = 
\cos\left(\frac{aq_\mu}{2}\right)\cos\left(\frac{aq_\nu}{2}\right)
      \left(\bar{\psi}\Dlr_\mu\Dlr_\nu \psi\right)^{(II)}(q) \,.
\end{equation}

We get the following results:

\vspace{0.5cm}
\underline{$\cO_{\{124\}}^{DD} \, (\tau_2^{(4)},C=-1)$}
\vspace{0.5cm}

In this case we have the mixing operators (\ref{mixing1}). The
corresponding $2 \times 2$-mixing matrices are
\begin{equation}
  \gamma=\left(
  \begin{array}{rr}
     \frac{25}{6} & -\frac{5}{6}\\
       0          & 0
  \end{array}
  \right) \,,
  \label{andim1}
\end{equation}
\begin{equation}
  B^{(0,I,II)} =\left(
  \begin{array}{rr}
       -11.56318    &  0.02414\\
         0          & 20.61780
  \end{array}
  \right) \,, 
\end{equation}
\begin{equation}
  B^{(1,I,II)} =\left(
  \begin{array}{rr}
       -2.89800    &  0.25529\\
         0          & -4.74556
  \end{array}
  \right) \,,
\end{equation}
\begin{equation}
  B^{(2,I,II)} =\left(
  \begin{array}{rr}
       -0.98387    &  -0.01557\\
         0          & -0.54317
  \end{array}
  \right)   \,.
\end{equation}
The matrix $B^{(I,II)}$ shows a rather small coefficient for the 
mixing between the operators
$\cO_{\{124\}}^{DD}$ and $\cO^{\partial\partial}_{\{124\}}$.
Thus it may be justified to neglect the mixing in practical applications,
where $\mu = 1/a$.

\vspace{0.5cm}
\underline{$\cO_1 \, (\tau_1^{(8)},C=-1)$}
\vspace{0.5cm}

The operators mixing with $\cO_1$ are given in (\ref{mixing2}). First we
consider the operators of the same dimension $\cO_1, \dots, \cO_7$.
To one-loop accuracy the operator $\cO_7$ does not contribute,
because its Born term vanishes, and 
we have to consider the following mixing set:
\begin{equation}
  \{\cO_1,\cO_2, \cO_3,\cO_4,\cO_5,\cO_6\} \,.
\end{equation}
The anomalous dimension matrix is
\begin{equation}
  \gamma =\left( \begin{array}{rrrrrr}
 \frac{25}{6}&-\frac{5}{6}& 0          & 0          & 0          & 0\\
    0        & 0          & 0          & 0          & 0          & 0\\
    0        & 0          & \frac{7}{6}&-\frac{5}{6}&1        &-\frac{3}{2}\\
    0        & 0          & 0          & 0          & 0          & 0 \\
    0        & 0          & 0          & 0          & 2          & -2 \\
    0        & 0          & 0          & 0          & -\frac{2}{3}& \frac{2}{3}
  \end{array}
  \right)
  \label{andim2}
\end{equation}
and the finite parts of the mixing matrix are given by (in doublets 
the upper number belongs to type I, the lower to type II of the 
realisation of the lattice covariant derivative)
\begin{equation}
  B^{(0,I,II)} =\left( \begin{array}{rrrrrr}
 -12.12740   & \left( \begin{array}{r}1.49127\\-2.73669 \end{array}\right)  
   & 0.36848       &
 \left( \begin{array}{r}-0.41595\\0.99336 \end{array}\right)       
   & 0.01562    & 0.14983\\[2ex]
    0       &  20.61780      & 0        & 0        & 0        & 0\\[2ex]
  3.30605    &  \left( \begin{array}{r}-8.01456\\18.18411 \end{array}\right) 
    &  -14.85157 & \left( \begin{array}{r}4.43061\\-4.30228 \end{array}\right) 
     & -0.92850   & 0.73802 \\[2ex]
 0       & 0            & 0           &  20.61780    & 0        & 0 \\
 0       & 3.26440       & 0           & 0           & 0.35008   & 0.01491 \\
 0       & 3.26440       & 0           & 0           & 0.00497    & 0.36003
  \end{array}
  \right) \,,
\end{equation}
\vspace*{0.5cm}
\begin{equation}
  B^{(1,I,II)} =\left( \begin{array}{rrrrrr}
  -2.92169   & \left( \begin{array}{r} 0.21269\\ 0.68643 \end{array}\right)  
  & 0.03276  & \left( \begin{array}{r} -0.01492\\ -0.17283 \end{array}\right)   
    & 0.01878   & -0.05696\\[2ex]
   0        &  -4.74556      & 0           & 0           & 0        & 0\\[2ex]
  -0.33335   &  \left( \begin{array}{r} 0.76570\\ 0.05510 \end{array}\right)  
  & -2.15228 & \left( \begin{array}{r} -1.20652\\ -0.96966 \end{array}\right)  
   & 1.75814   & -2.29837 \\[2ex]
 0    & 0            & 0           & -4.74556       & 0        & 0 \\
 0    & 1.44106       & 0           & 0           & -1.64790    & -0.86576 \\
 0    & 1.44106       & 0           & 0           & -0.28859    & -2.22507
  \end{array} 
  \right) \,,
\end{equation}
\vspace*{0.5cm}
\begin{equation}
  B^{(2,I,II)} =\left( \begin{array}{rrrrrr}
 -0.98166   & \left( \begin{array}{r}-0.07815\\-0.10117 \end{array}\right)  
   & -0.02914 & \left( \begin{array}{r}0.03475\\0.04243 \end{array}\right)
   & -0.00999   & 0.00688\\[2ex]
 0      & -0.54317       & 0           & 0           & 0        & 0\\[2ex]
 0.37050    & \left( \begin{array}{r}-0.55068\\0.21545 \end{array}\right)   
   & -1.70741  & \left( \begin{array}{r}0.37132\\0.11594 \end{array}\right) 
   & -0.44295   & 0.10328 \\[2ex]
 0   & 0            & 0           & -0.54317      & 0        & 0 \\
 0   & 1.41570        & 0           & 0           & -1.70334   & 0.56763 \\
 0   & 1.41570       & 0           & 0           & 0.18921    & -1.32493
  \end{array}
  \right) \,.
\end{equation}
The matrices $B^{(k,I,II)}$ show sizeable coefficients for the 
mixing of $\cO_1$ with other operators, especially with $\cO_2$
containing two external ordinary derivatives.

There is also a possible mixing between $\cO_1$
and the lower-dimensional operator $\cO_8$ in (\ref{mixing2}).
Indeed, we find in the one-loop approximation that the
vertex function of $\cO_1$ contains a term $\propto 1/a$:
\begin{equation}
  \cO_1\bigg|_{\frac{1}{a}-{\rm part}} =
  \gRcf (-0.51771-0.08325\,c_{sw} - 0.00983\,c_{sw}^2)\,
                                 \frac{1}{a}\, \cO_8^{\rm Born}\,.
  \label{Ba1}
\end{equation}
This mixing leads to a contribution which diverges like
the inverse lattice spacing in the continuum limit.
Thus the perturbative calculation of the mixing coefficient is not
reliable and the operator $\cO_8$ has to be subtracted non-perturbatively
from the operator $\cO_1$.

\vspace{0.5cm}
\underline{$\cO^{5,DD}_{\{124\}} \, (\tau_3^{(4)},C=+1)$}
\vspace{0.5cm}

In this case we have to consider the operators in (\ref{mixing3}).
The anomalous dimension matrix is given by (\ref{andim1}), the finite
contributions are collected in the matrices
\begin{equation}
  B^{(0,I,II)} =\left( \begin{array}{rr}
    -12.11715     & 0.16673\\
      0          & 15.79628
  \end{array} \right) \,,
\end{equation}
\begin{equation}
   B^{(1,I,II)} =\left( \begin{array}{rr}
    -1.51925     & 0.00505\\
      0          & 0.24783
  \end{array} \right) \,,
\end{equation}
\begin{equation}
   B^{(2,I,II)} =\left( \begin{array}{rr}
    -1.71846     & 0.00711\\
      0          & -2.25137
  \end{array}  \right) \,.
\end{equation}
As in the case of the operator $\cO_{\{124\}}^{DD}$
above, the mixing coefficient is rather small.

\vspace{0.5cm}
\underline{$\cO_1^5 \, ( \tau_2^{(8)},C=+1)$}
\vspace{0.5cm}

First we discuss the mixing of operators of the same dimension 
in (\ref{mixing4}). The set of contributing operators
is found to be
\begin{equation}
  \{\cO^5_1,\cO^5_2,\cO^5_3,\cO^5_4,\cO^5_5,\cO^5_6 \} \,.
\end{equation}
As in the case of the operator $\cO_1$, one operator -- here $\cO^5_7$ --
does not contribute to mixing in one-loop order.
The finite contributions are
\begin{equation}
  B^{(0,I,II)} =\left( \begin{array}{rrrrrr}
  -12.86094 & \left( \begin{array}{r}-2.06532\\1.48943\end{array}\right) 
  &  0.34900 & \left( \begin{array}{r}0.85381 \\-0.33110 \end{array} \right)  
  & 0.05113  & 0.05942\\[2ex]
    0     &  15.79628      & 0           & 0           & 0       & 0\\[2ex]
  3.42196  &  \left( \begin{array}{r}15.82073\\-7.30020\end{array} \right)  
  & -15.35920 & \left( \begin{array}{r}-5.16392 \\2.54306\end{array}\right)   
  &  0.17014 & -0.94314 \\[2ex]
    0  & 0            & 0           & 15.79628   & 0       & 0 \\
    0  & -8.91237      & 0           & 0           & 0.95969  & -0.95969 \\
    0  & -8.91237      & 0           & 0           & -0.31990  & 0.31990
  \end{array}
  \right) \,,
\end{equation}
\begin{equation}
  B^{(1,I,II)} =\left( \begin{array}{rrrrrr}
  -1.49316  & \left( \begin{array}{r} 0.03027\\ -0.17480\end{array}\right) 
  &  0.00750 & \left( \begin{array}{r} -0.03022 \\ 0.09858 \end{array} \right) 
  & -0.01290  & 0.03622\\[2ex]
  0     &  0.24783      & 0           & 0           & 0       & 0\\[2ex]
  0.09099 & \left( \begin{array}{r} -0.99704\\ 1.07971\end{array} \right)   
  & -2.30129 & \left( \begin{array}{r} 1.25511 \\ 0.56286\end{array}\right) 
  & -0.13420 & -0.48232 \\[2ex]
  0  & 0            & 0           & 0.24783     & 0       & 0 \\
  0  & 13.26724      & 0           & 0           & -3.27954  & 1.78029 \\
  0  & 13.26724      & 0           & 0           &  0.59343  & -2.09268
  \end{array}
  \right) \,,
\end{equation}
\begin{equation}
  B^{(2,I,II)} =\left( \begin{array}{rrrrrr}
  -1.68673 & \left( \begin{array}{r}0.03501\\0.19704\end{array}\right) 
  &  -0.00612 & \left( \begin{array}{r}-0.01275 \\-0.06677 \end{array} \right) 
  & 0.01000  & -0.00692\\[2ex]
  0     &  -2.25137      & 0           & 0           & 0       & 0\\[2ex]
  0.16581 &  \left( \begin{array}{r}0.28263\\0.13256\end{array} \right)   
  & -1.36546 & \left( \begin{array}{r}-0.30614 \\-0.25600\end{array}\right)  
  &  0.44299 & -0.10322 \\[2ex]
  0  & 0            & 0           & -2.25137     & 0       & 0 \\
  0  & -1.41570      & 0           & 0           & -0.97445  & -0.88857 \\
  0  & -1.41570      & 0           & 0           & -0.29619  & -1.56683
  \end{array}
  \right) \,.
\end{equation}
The anomalous dimension matrix is the same as for the operators without
$\gamma_5$, see (\ref{andim2}). Again, some of the mixing coefficients
are non-negligible. The mixing between $\cO^5_1$ and $\cO^5_3$
is present also in the forward case.

We also find mixing with a lower-dimensional operator, in this case it
is the operator $\cO_8^5$ in (\ref{mixing4}). The corresponding contribution in
the vertex function of $\cO^5_1$ reads
\begin{equation}
  \cO_1^5\bigg|_{\frac{1}{a}-{\rm part}} =
  -\gRcf \,(0.25231 - 0.02507 \, c_{sw} +0.01046 \, c_{sw}^2)\,
                              \frac{1}{a}\,\cO_8^{5,{\rm Born}} \,.
   \label{Ba2}
\end{equation}

\vspace{0.5cm}
\underline{$\cO^T_1 \, (\tau_2^{(3)},\, C=-1)$}
\vspace{0.5cm}

For the operators  (\ref{mixing5}) the matrix of anomalous dimensions 
is given by
\begin{equation}
  \gamma=\left( \begin{array}{rr}
    \frac{13}{3} & -\frac{2}{3}\\
      0          & 1
  \end{array}
  \right) \,,
\label{andimT1}
\end{equation}
and the finite contributions are
\begin{equation}
  B^{(0,I,II)} = \left( \begin{array}{rr}
    -11.54826     & 0.21894 \\
      0          & 17.01808
  \end{array}  \right) \,, 
\end{equation}
\begin{equation}
 B^{(1,I,II)} = \left( \begin{array}{rr}
    -2.41077     & -0.05383 \\
      0          & -3.91333
  \end{array}  \right) \,,
\end{equation}
\begin{equation}
  B^{(2,I,II)} =\left( \begin{array}{rr}
    -1.51175     & -0.00614 \\
      0          & -1.97230
  \end{array}
  \right) \,.
\end{equation}

\vspace{0.5cm}
\underline{$\cO^T_3 \, (\tau_3^{(3)},\, C=-1)$}
\vspace{0.5cm}

The operators (\ref{mixing6}) have the same anomalous dimension matrix
(\ref{andimT1}) as the previous case. For the finite pieces we obtain
\begin{equation}
  B^{(0,I,II)} = \left( \begin{array}{rr}
  -11.86877     & 0.27533 \\
    0          & 17.01808
  \end{array} \right) \,, 
\end{equation}
\begin{equation}
  B^{(1,I,II)} = \left( \begin{array}{rr}
  -2.30651     & -0.01831 \\
    0          & -3.91333
  \end{array} \right) \,,
\end{equation}
\begin{equation}
  B^{(2,I,II)} = \left( \begin{array}{rr}
  -1.34908     & 0.01726 \\
    0          & -1.97230
  \end{array} \right) \,.
\end{equation}

\vspace{0.5cm}
\underline{$\cO^T_5 \, (\tau_2^{(6)},\, C=-1)$}
\vspace{0.5cm}

We find for the case (\ref{mixing7}) the finite mixing contributions
\begin{equation}
  B^{(0,I,II)} = \left( \begin{array}{rr}
  -11.74773     & 0.23797 \\
    0          & 17.01808
  \end{array}  \right) \,,
\end{equation}
\begin{equation}
   B^{(1,I,II)} = \left( \begin{array}{rr}
  -2.36201     & -0.04490 \\
    0          & -3.91333
  \end{array}  \right) \,,
\end{equation}
\begin{equation}
   B^{(2,I,II)} = \left( \begin{array}{rr}
  -1.45084     & 0.00898 \\
    0          & -1.97230
  \end{array}  \right)
\end{equation}
with the anomalous dimension matrix (\ref{andimT1}).

\section{Tadpole improvement}

It is well known that many results of (naive) lattice perturbation 
theory do not agree very well with their counterparts determined 
from Monte Carlo calculations. One main reason for these 
discrepancies is the appearance of gluon tadpoles, which are
typical lattice artefacts. They turn the bare coupling $g$ into a poor
expansion parameter. As a remedy the so-called tadpole 
(or mean field) improvement 
has been proposed~\cite{Lepage:1992xa}, a rearrangement of the 
perturbative series making use of the variable $u_0$, the fourth root
of the measured value of the plaquette,
\begin{equation}
u_0 = \langle\frac{1}{N_c} {\rm Tr}U_\Box\rangle^{\frac{1}{4}} \,.
\end{equation}
Its value depends on the coupling $g^2=6/\beta$ where it has been measured.

In case of mixing the tadpole improvement works as follows.
Scaling the link variables $U_\mu$ with $u_0$
\begin{equation}
U_\mu(x) = u_0\,\left( \frac{U_\mu(x)}{u_0} \right) = u_0\,\overline{U}_\mu(x)
\end{equation}
one finds an expression for the vertex function $\Gamma_j$ of
an operator $\cO_j$ containing $n_j$ covariant derivatives 
which is of the form
\begin{equation}
\Gamma_j (U_\mu(x)) = 
   u_0^{n_j} \Gamma_j (\overline{U}_\mu(x))\,.
\end{equation}
$\Gamma_j (\overline{U}_\mu(x)) = 
                 u_0^{-n_j} \Gamma_j (U_\mu(x))$ 
is expected to have a better converging perturbative expansion,
which is obtained by inserting the expansions of $u_0^{-n_j}$ and 
$\Gamma_j (U_\mu(x))$.  
To one-loop accuracy, $u_0$ is given by  
\begin{equation}
u_0 = 1 - \frac{g^2 C_F}{16\,\pi^2} \, \pi^2 +O(g^4) \,.
\end{equation}
Note that at the one-loop level we do not have to distinguish between
$g$ and $\gR$ in the perturbative expressions.
The exponent of $u_0$ depends on $j$ because in general the 
mixing operators have different numbers of covariant derivatives 
(see, e.g., (\ref{mixing1}) or (\ref{mixing2})). An external ordinary 
derivative ($\partial$) does not provide a factor of $u_0$. 
Taking into account the mean field value for the wave function
renormalisation constant for massless Wilson and clover fermions
\begin{equation}
Z_{\psi}^{MF} = u_0
\end{equation}
the mean field $Z$ factor for each operator $\cO_j$ contributing to the
mixing reads 
\begin{equation}
Z_j^{MF} = u_0^{1-n_j} \,,
\end{equation}
and the tadpole improved matrix of renormalisation constants
is given by
\begin{eqnarray}
Z_{ij}^{TI} = 
Z_{ij} \frac{Z_j^{MF}}{Z_j^{MF, \mathrm {pert}}} =
  u_0^{1-n_j}\,\left(1- \frac{g^2 C_F}{16\,\pi^2}(n_j-1)\,\pi^2
         + O(g^4) \right)\,Z_{ij}\,.
\label{TI2}
\end{eqnarray}
Additionally, we replace the parameters
$g$ and $c_{sw}$ by their ``boosted'' counterparts
\begin{eqnarray}
g_{TI}^2 \equiv g^2 \, u_0^{-4}\,, \quad
c^{TI}_{sw} \equiv c_{sw} \, u_0^{3}\,.
\label{TI3}
\end{eqnarray}
Combining (\ref{Zjk}), (\ref{TI2}) and (\ref{TI3}) we obtain 
for the tadpole improved matrix in one-loop order
\begin{eqnarray}
Z_{ij}^{TI,(m)} &=& u_0^{1-n_j}\, \left(\delta_{ij} -
  \gTIcf \left(\gamma_{ij}\,\ln(a^2\mu^2)
  + B_{ij}^{(m)}(c^{TI}_{sw}) + (n_j-1)\pi^2\, \delta_{ij}\right)\right)
  \nonumber
  \\
  &\equiv& u_0^{1-n_j}\, \left(\delta_{ij} -
  \gTIcf \left(\gamma_{ij}\,\ln(a^2\mu^2)
  + B_{ij}^{TI,(m)}(c^{TI}_{sw})\right)\right) \,.
\label{TI4}
\end{eqnarray}

Let us exemplify the impact of tadpole improvement in some typical
cases. We choose $\mu=1/a$, $c_{sw}=1+O(g^2)$
and $u_0=0.8778$ corresponding to quenched calculations at $\beta=6$.
For operators with one covariant derivative the tadpole improvement 
procedure is rather simple. Because $n_1 = 1$ the only effect 
consists in replacing in (\ref{Z}) and (\ref{tabZ}) $c_{sw}$ 
by $c^{TI}_{sw}$ and $g$ by $g_{TI}$. For the operators 
belonging to the representation $\tau_3^{(6)}$, e.g., we get
\begin{equation}
Z = 1.028  \quad \rightarrow \quad Z^{TI} = 1.023\,.
\end{equation}

The operators for the second moments of GPDs are a bit more involved.
First we consider the simple mixing 
$\cO_{\{124\}}^{5,DD} \leftrightarrow \cO_{\{124\}}^{5,\partial\partial}$
(\ref{mixing3}). Without tadpole improvement we 
obtain the mixing matrix of renormalisation constants as
\begin{equation}
  Z =\left( \begin{array}{rr}
  1.12965        &   -0.00151 \\
    0            &    0.88354
  \end{array}
  \right) \,,
\end{equation}
The tadpole improved result is
\begin{equation}
  Z^{TI} =\left( \begin{array}{rr}
  1.20501        &   -0.00216   \\
    0            &    0.81458
  \end{array}
  \right) \,,
\end{equation}

It is instructive to compare the one-loop corrections
for the renormalisation constants, i.e.\ $B^{(m)}_{ij}(c_{sw})$ for the
unimproved case (\ref{Zjk}) and $B_{ij}^{TI,(m)}(c^{TI}_{sw})$
for the tadpole improved case (\ref{TI4}). With the parameters
given above we get for the operators (\ref{mixing3})
\begin{equation}
  B =\left( \begin{array}{rr}
  -15.35486     &    0.17889    \\
    0            &  13.79274
  \end{array}
  \right)\,,
  \label{BUI1}
\end{equation}
and
\begin{equation}
  B^{TI} =\left( \begin{array}{rr}
  -4.06129       &   0.17340     \\
    0          &     5.06434
  \end{array}
  \right) \,.
  \label{BTI1}
\end{equation}
Equations (\ref{BUI1}) and (\ref{BTI1}) show that the one-loop 
corrections on the diagonal have been reduced significantly. This is
in accordance with the aims of tadpole improvement.

The same procedure can be applied to the more complicated 
set $\mathcal{O}_1, \dots, \mathcal{O}_6$ $(\tau_1^{(8)},C=-1)$ 
from (\ref{mixing2}). We obtain for the unimproved case
the mixing matrix of renormalisation constants (lattice covariant 
derivative type $m=I$) as
\begin{equation}
  Z^{(I)} =\left( \begin{array}{rrrrrr}
  1.13535       &   -0.013727     & -0.00314 &   0.00334   & -0.00021  
                                                                & -0.00084\\
    0          &    0.87057     & 0      & 0               & 0         & 0\\
  -0.02823       &  0.06585     & 1.15799  & -0.03036      & -0.00327   
                                                                & 0.01230 \\
    0          & 0            & 0           & 0.87057      & 0         & 0 \\
    0          & -0.05168       & 0           & 0          & 1.02534   
                                                                & 0.00239 \\
    0          & -0.05168       & 0           & 0          & 0.00080   
                                                                & 1.02693
  \end{array}
  \right) \,,
\end{equation}
The tadpole improved result reads
\begin{equation}
  Z^{TI,(I)} =\left( \begin{array}{rrrrrr}
  1.21508        &   -0.01997     & -0.00611 &   0.00512    & -0.00034   
                                                                & -0.00163 \\
    0          &    0.78680     & 0      & 0           & 0        & 0\\
  -0.05265       &    0.09673     & 1.25616  & -0.04724      & -0.00082 
                                                                & 0.01094 \\
    0          & 0            & 0           & 0.78680  & 0        & 0 \\
    0          & -0.06100       & 0           & 0      & 1.02195    
                                                                & 0.00442 \\
    0          & -0.06100       & 0           & 0      & 0.00147   & 1.02490
  \end{array}
  \right) \,.
\end{equation}
The renormalisation and mixing matrices for all other cases can be 
treated analogously. They show a similar behaviour: The $2\times 2$ 
matrices exhibit rather small mixing coefficients, the other $6\times 6$
problem (\ref{mixing4}) is almost identical to the example discussed above.

\section{Summary}

In this paper we have considered quark-antiquark operators needed for
the computation of the first two moments of GPDs and meson distribution 
amplitudes within the framework of lattice QCD. In one-loop lattice 
perturbation theory we have calculated the non-forward quark matrix elements 
of these operators employing clover improved Wilson fermions and 
Wilson's plaquette action for the gauge fields. From the results 
we have evaluated the matrices of renormalisation and mixing 
coefficients in the $\MS$-scheme. 

For the operators with only one derivative (relevant for the first
moments) we could take over the numbers obtained for 
the first moments of ordinary parton distributions. 
The results for the second moments
generalise the numbers calculated with Wilson fermions~\cite{Gockeler:2004xb}.
The general conclusions concerning the mixing properties
remain unchanged. 

If there is only mixing between one operator with two 
covariant derivatives $D$ and one operator with two external 
derivatives $\partial$ the mixing coefficient turns out to be 
rather small. In the two cases (\ref{mixing2}) and (\ref{mixing4}) 
with eight potentially mixing operators the mixing effects are more 
severe. Moreover, taking  $\cO_1$ from (\ref{mixing2})
or $\cO^5_1$ from (\ref{mixing4}) as the operator to be
measured in a numerical simulation we find in each case mixing 
with a lower-dimensional operator (cf.\ (\ref{Ba1}) and (\ref{Ba2})). 
This could lead to difficulties, because $1/a$ effects are hard 
to get under control. For overlap fermions, however, these mixings
with lower-dimensional operators of different chirality would be absent. 

Additionally, we have discussed tadpole improvement
with special attention to mixing operators. We have given the 
general prescription for mean field improvement and have shown 
how it works for selected examples.

\section*{Acknowledgements}
This work
is supported by DFG under contract FOR 465 (Forschergruppe
Gitter-Hadronen-Ph\"{a}nomeno\-logie) and by the EU
Integrated Infrastructure Initiative Hadron Physics under contract
number RII3-CT-2004-506078.

\end{document}